# The effect of surface quenching coefficients of $O_2(a^1\Delta g)$ and $O_2(b^1\Sigma g^+)$ on capacitively coupled Ar/O₂ discharge: A global/equivalent circuit model study

Wan Dong, Yi Wang, Yi-Fan Zhang, Yuan-Hong Song[*]

Key Laboratory of Materials Modification by Laser, Ion and Electron Beams (Ministry of Education), School of Physics, Dalian University of Technology, Dalian 116024, People's Republic of China

[*]E-mail: songyh@dlut.edu.cn

**Abstract**

Capacitively coupled discharges operated in mixtures of Ar and O₂ are extensively utilized in plasma etching and deposition processes due to the oxidative properties and precursor functionality of the reactive species produced in the discharge. In Ar/O₂ discharges, the surface quenching coefficient of $O_2(a^1\Delta g)$ is known to affect this metastable density, which, in turn, affects the electronegativity and other important plasma characteristics. In this work, in addition to $O_2(a^1\Delta g)$, $O_2(b^1\Sigma g^+)$ and its associated reactions are incorporated into a global/equivalent circuit model of an Ar/O₂ discharge. By independently adjusting the quenching coefficients of both metastable species, changes of these surface coefficients are found to significantly affect the discharge characteristics, indicating that the role of $O_2(b^1\Sigma g^+)$ cannot be neglected. The effects of their respective surface quenching coefficients of these metastables based on various wall materials on the discharge are revealed including their effects on different particle species densities, plasma impedance, voltage drops across the sheaths, as well as plasma power absorption.

**Keywords**: Capacitive Ar/O₂ plasmas, global/equivalent circuit model, quenching coefficients.

## 1 Introduction

Radio-frequency capacitively coupled plasmas (RF CCPs) are extensively employed in plasma-enhanced etching and deposition processes [1-8]. Gas mixtures containing $O_2$ combined with $CF_4$ or $C_4F_8$ and Ar are commonly used in plasma etching applications due to their effectiveness in controlling etching rates, selectivity, and profile shaping [9-12]. Numerous studies have investigated RF CCPs operated in pure oxygen and in gas mixtures containing $O_2$ to understand the discharge characteristics and to control these discharges. These studies primarily focus on discharge modes [13-15], heating mechanisms [16-18], and the characteristics of etching and deposition processes [19-21].

Moreover, the evolution of key species involved in industrial processes [22-24], such as $O(^1D)$ and $O(^3P)$, as well as particles that are highly relevant to the properties of oxygen discharges [25-28], has been investigated as a function of external control parameters. In particular, $O_2(a^1\Delta g)$ has attracted considerable attention due to its critical role for plasma behavior. Its surface quenching coefficient determines the loss of $O_2(a^1\Delta g)$ at the boundaries, which further affects its density. Besides, $O_2(a^1\Delta g)$ can react with $O^-$, leading to a reduction in $O^-$ density via electron detachment, significantly influencing the plasma electronegativity, effective electron temperature, and the electron power absorption dynamics [25-28]. For example, in capacitively coupled oxygen discharges using a one-dimensional Particle-in-Cell/Monte Carlo Collision (PIC/MCC), different electron heating mechanisms are observed depending on whether $O_2(a^1\Delta g)$ is included in the $O_2$ reaction set [25]. When $O_2(a^1\Delta g)$ is excluded, electron heating is jointly dominated by bulk and sheath heating. In contrast, including $O_2(a^1\Delta g)$ shifts the heating mechanism to be primarily dominated by sheath heating. In addition, Proto et al. [26] investigated the impact of the surface quenching coefficient of $O_2(a^1\Delta g)$ on the electron energy probability function (EEPF) in single frequency capacitively coupled oxygen discharges. Their results revealed that, increasing the quenching coefficient causes the EEPF to gradually shift from a concave (bi-Maxwellian) to a convex profile. This will significantly affect the plasma discharge and particle generation.

Gibson et al. [27] investigated the influence of $O_2(a^1\Delta g)$ on charged species behaviour in capacitively coupled oxygen plasmas using a one-dimensional fluid simulation. The results indicate that, when the density of $O_2(a^1\Delta g)$ reaches 16% of the background gas $O_2$ density, the charged particle distribution shows similar spatio-temporally averaged densities of electrons and negative ions, with a space-time averaged electronegativity of approximately 1. In contrast, when the $O_2(a^1\Delta g)$ density is only 0.5% of the $O_2$ density, the electron density is significantly lower than that of the negative ions, resulting in a substantially higher space-time averaged electronegativity of about 17. These studies highlight the crucial role of $O_2(a^1\Delta g)$ in oxygen plasma discharges.

Additionally, in the $O_2$ reaction set, similar to $O_2(a^1\Delta g)$, electrons can also collide with $O_2$

to generate $O_2(b^1\Sigma_g^+)$ via electron impact excitation. The electron energy excitation thresholds of $O_2(a^1\Delta_g)$ and $O_2(b^1\Sigma_g^+)$ are 0.98 eV and 1.627 eV, respectively. These processes have large electron-impact cross sections, resulting in high excitation rates and densities of $O_2(a^1\Delta_g)$ and $O_2(b^1\Sigma_g^+)$. However, in simulation studies of capacitively coupled $O_2$ and $Ar/O_2$ discharges, $O_2(b^1\Sigma_g^+)$ is often neglected. It has often been assumed in previous studies that the density of $O_2(b^1\Sigma_g^+)$ is low [29, 30]. The reason for this assumption is that the primary loss mechanism for both $O_2(a^1\Delta_g)$ and $O_2(b^1\Sigma_g^+)$ is quenching to the wall, and the quenching coefficient of $O_2(b^1\Sigma_g^+)$ is significantly higher than that of $O_2(a^1\Delta_g)$ [31, 32]. However, Proto et al. [26] found that the wall quenching coefficient of $O_2(b^1\Sigma_g^+)$ may have been overestimated. In addition, reactions generating $O_2(b^1\Sigma_g^+)$ have been neglected, such as the reaction $O^*+O_2 \rightarrow O+O_2(b^1\Sigma_g^+)$ [31-33]. As a result, the density of $O_2(b^1\Sigma_g^+)$ has likely been underestimated in previous studies. For ICP oxygen discharges, within the explored pressure range (< 100 mTorr), both $O_2(a^1\Delta_g)$ and $O_2(b^1\Sigma_g^+)$ are present in significant densities, with the density of $O_2(b^1\Sigma_g^+)$ even being higher than that of $O_2(a^1\Delta_g)$. The study also found that $O_2(A^3\Sigma_u^+, A^3\Delta_u, c^1\Sigma_u^-)$ contributes to the production of $O_2(b^1\Sigma_g^+)$ [34]. Recent simulation studies have also highlighted the importance of $O_2(b^1\Sigma_g^+)$ in RF capacitively coupled oxygen discharges, and have compared its role with that of $O_2(a^1\Delta_g)$. These studies revealed that $O_2(b^1\Sigma_g^+)$ has a significant impact on electronegativity, electron temperature, and discharge modes [35, 36].

In previous studies of capacitively coupled oxygen discharges, attention has primarily been paid to the effects of the surface quenching coefficients of $O_2(a^1\Delta_g)$ and $O_2(b^1\Sigma_g^+)$ on charged particle densities, while changes in neutral densities and the key mechanisms of particle generation and loss have not been considered and some relevant collision reactions have not been taken into account. Furthermore, current research on the effects of the surface quenching coefficients of $O_2(a^1\Delta_g)$ and $O_2(b^1\Sigma_g^+)$ has been mainly focused on pure $O_2$ discharges rather than $Ar/O_2$ discharges generally used in industry. In this work, the effects of the surface quenching coefficients of $O_2(a^1\Delta_g)$ and $O_2(b^1\Sigma_g^+)$ on the characteristics of capacitively coupled $Ar/O_2$ gas mixture discharges and the relevant collision reaction channels will be systematically investigated based on an extended $Ar/O_2$ reaction set [37] in the frame of a global/equivalent circuit model, aiming to provide a valuable reference for future simulation research.

**2 Model description**

To realize simulations of RF CCP discharges operated in complex chemistries in this work, a nonlinear equivalent circuit model is coupled to a global model of the plasma. In the global model, the calculation of the electron density, electron temperature and ion densities rely on the energy balance equation, the particle number conservation equations for each species and the quasi-neutrality condition [29, 37-40]. In the equivalent circuit model, the plasma is represented as a combination of equivalent LCR elements [37, 40, 41, 42]. It is assumed that the plasma between the two electrodes can be divided into a bulk region and two sheath regions. The plasma bulk is quasi-neutral with uniform particle densities. The sheaths near the powered and grounded electrode are modeled as a parallel circuit consisting of a constant current source, a diode, and a nonlinear capacitor. The bulk region can be represented by a series of a resistor and an inductor. According to Kirchoff's laws, the relationships among the sheath voltage drop at the driven electrode, $V_{s,p}$, and grounded electrode, $V_{s,g}$, the radio-frequency plasma current $I_{rf}$, and the bias sheath capacitor voltage $V_{C_B}$ can be obtained from a set of time-dependent nonlinear ordinary differential equations for a fixed set of circuit elements, which depend on plasma parameters such as the electron density. A detailed description of this model as well as specific model and boundary conditions can be found in our previous work [37, 40].

The coupling between the global model and the nonlinear equivalent circuit model is realized as follows: First, based on an initial set of plasma parameters, including estimated values for each component in the equivalent circuit, a transient simulation is conducted using the equivalent circuit model to obtain the current and voltage at each branch and node. From these results, the plasma's absorbed power and sheath voltage drops can be calculated. These values are then fed into the global model to update the electron density, ion density, and electron temperature. The updated plasma parameters are subsequently input into the nonlinear equivalent circuit model to calculate the specific values of the circuit elements and recalculate the absorbed power and sheath voltage drops, which are again provided to the global model. This iterative process is continued until all physical quantities reach convergence.

For the collision reaction set of Ar/O$_2$ used in this work, 15 species and 102 reactions are considered. These species include electrons, O$_2$, Ar, O$_2$(a$^1\Delta$g), O$_2$(b$^1\Sigma$g$^+$), O$_2$(A$^3\Sigma_u^+$, A$^3\Delta_u$, c$^1\Sigma_u^-$), O$_3$, O, O$^*$, O$_2^+$, O$^+$, O$_2^-$, O$^-$, Ar$^*$ (specifically the 1$_{s5}$ and 1$_{s3}$ levels), Ar$^+$. The list of reactions [29, 34, 43-45] is shown in the Appendix table A1 and A2. And, the calculated results from this model show excellent agreement with those of Ref. [29], with charged species densities matching in both trend and magnitude. These to some extent validate the reliability of the model and the collision reaction set used in this work.

In this work, a capacitively coupled Ar/O$_2$ (50/50) mixture discharge is studied, with an electrode gap of 4.5 cm, a driving frequency of 13.56 MHz, and a voltage amplitude of 150 V. This study investigates the effects of gas pressure (20-100 mTorr) and the surface quenching coefficients of O$_2$(a$^1\Delta$g), and O$_2$(b$^1\Sigma$g$^+$) based on different chamber wall materials on species densities and power absorption. Table 1 lists the quenching coefficients for O$_2$(a$^1\Delta$g) ($\gamma$a), O$_2$(b$^1\Sigma$g$^+$) ($\gamma$b), corresponding to three specific wall materials: Teflon, Copper (Cu), and stainless steel. From table 1, $\gamma$b is consistently higher than $\gamma$a for these three materials.

Table 1 The quenching coefficients of the O$_2$(a$^1\Delta$g) ($\gamma$a) and O$_2$(b$^1\Sigma$g$^+$) ($\gamma$b) under different chamber materials [18, 46-48].

| # | γa | γb |
| --- | --- | --- |
| Teflon | $3\times10^{-3}$ [46] | $4.5\times10^{-3}$ [47] |
| Cu | $2.9\times10^{-4}$ [48] | $1\times10^{-2}$ [47] |
| Stainless steel | $7\times10^{-2}$ [18] | $1\times10^{-1}$ [18] |

## 3 Results

This study presents a systematic investigation of capacitively coupled $Ar/O_2$ mixed-gas discharges under varying gas pressures and quenching coefficients of $O_2(a^1\Delta g)$, and $O_2(b^1\Sigma g^+)$ based on different chamber wall materials. To emphasize the critical role of the quenching coefficients of $O_2(a^1\Delta g)$, and $O_2(b^1\Sigma g^+)$ in determining electron temperature, electronegativity, and the densities of charged and neutral species, in section 3.1, γa (or γb) is fixed while varying another coefficient. Section 3.2 examines the impact of quenching coefficients for both $O_2(a^1\Delta g)$ and $O_2(b^1\Sigma g^+)$ depending on different materials, on particle densities and plasma characteristics.

### 3.1 Impact of γa or γb on Ar/O₂ discharge characteristics

In this section, a capacitively coupled $Ar/O_2$ (50/50) mixture discharge is studied, featuring an electrode gap of 4.5 cm and operated at a pressure of 100 mTorr, a frequency of 13.56 MHz, and a voltage amplitude of 150 V. For $O_2(a^1\Delta g)$, γa ranges from approximately $1.3\times10^{-5}$ for glass electrodes to about $1\times10^{-2}$ for Pt or Ag electrodes [43]. For $O_2(b^1\Sigma g^+)$, γb also reaches its minimum (around $2\times10^{-3}$) for glass electrodes, while its maximum value is approximately 0.1 [18, 44]. Therefore, in this section, the value of γa is set in the range of 0.00001 to 0.01, while γb ranges approximately from 0.001 to 0.1, covering the same material ranges. Besides, when examining the effect of γa on the discharge, γb is held constant at 0.1. And, when the γb is varied, γa is fixed at 0.007.

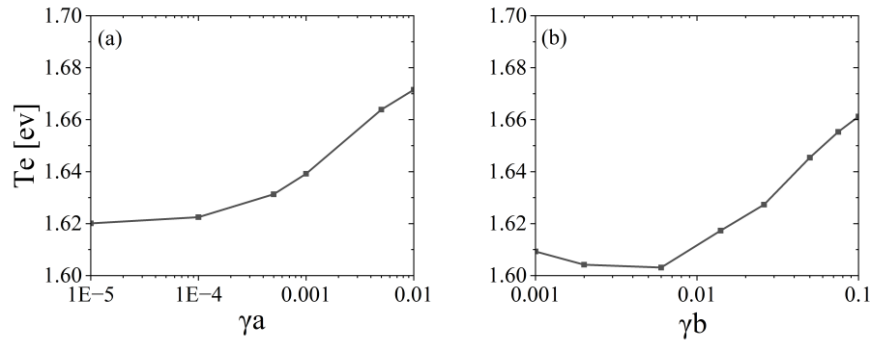

Figure 1 Electron temperature as a function of γa (a), and γb (b). Discharge conditions: gas ratio of $Ar/O_2$ is 50/50, with an electrode gap of 4.5 cm, a fixed pressure of 100 mTorr, a frequency of 13.56 MHz, and a voltage amplitude of 150 V.

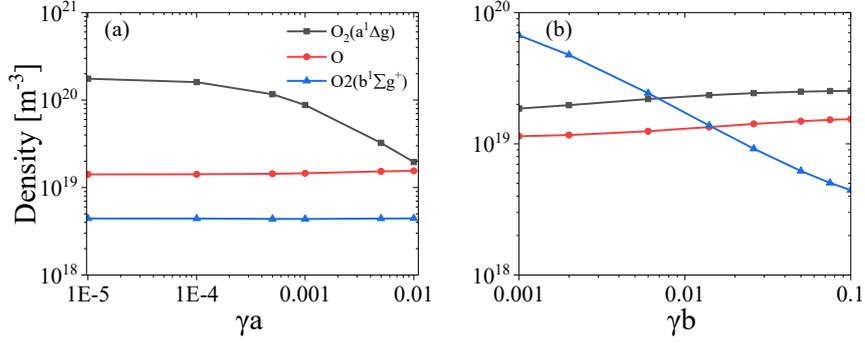

Figure 2 Densities of O (red line), $O_2(a^1\Delta g)$ (black line), $O_2(b^1\Sigma g^+)$ (blue line) as a function of $\gamma a$ (a), and $\gamma b$ (b). The discharge conditions are the same as those in figure 2

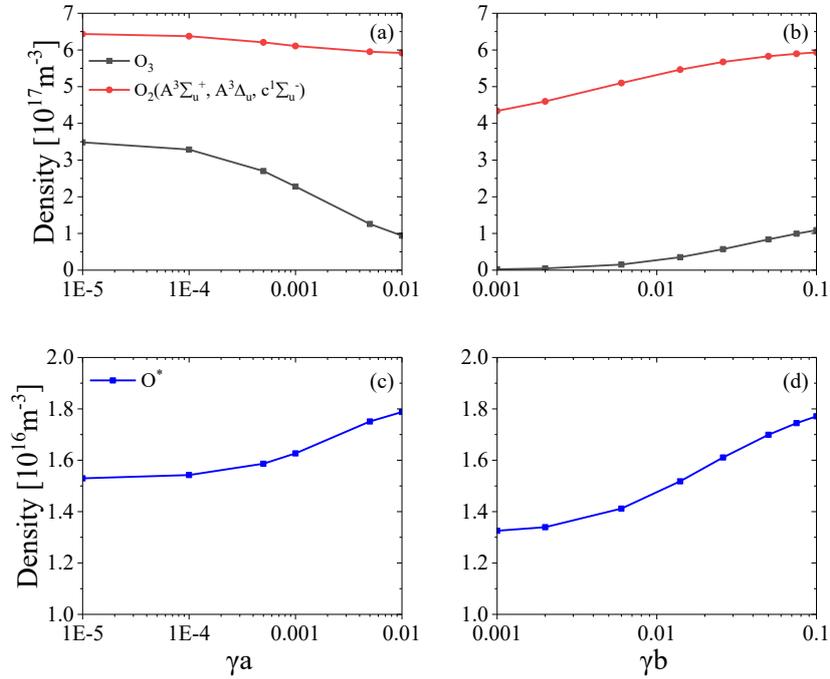

Figure 3 Densities of $O_3$, $O_2(A^3\Sigma_u^+, A^3\Delta_u, c^1\Sigma_u^-)$, $O^*$ as a function of $\gamma a$ (a), (c), and $\gamma b$ (b), (d). The discharge conditions are the same as those in figure 2

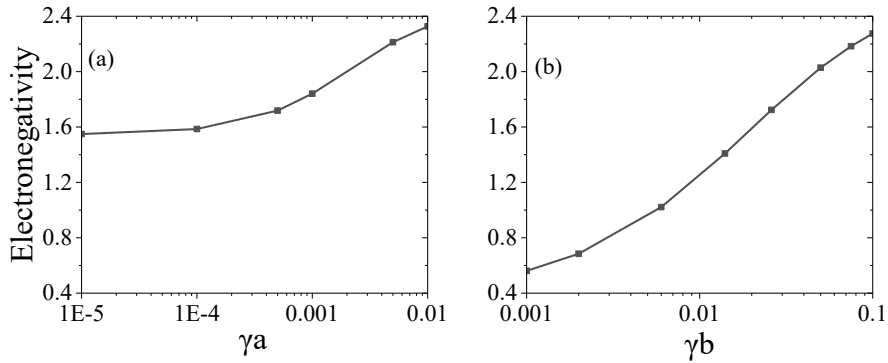

Figure 4 Electronegativity as a function of $\gamma a$ (a), and $\gamma b$ (b). The discharge conditions are the same as those in figure 2.

Figure 1 shows the electron temperature for different values of γa and γb, respectively. From figure 1, the electron temperature initially remains nearly constant and then exhibits a slight increase as the quenching coefficient enhances. A minor difference is observed when γb increases from 0.001 to 0.006, during which the electron temperature experiences a slight decrease, but this variation is less than one percent. Figures 2 and 3 illustrate the changes in the densities of O, $O_2(a^1\Delta g)$, $O_2(b^1\Sigma g^+)$, $O_3$, $O_2(A^3\Sigma_u^+, A^3\Delta_u, c^1\Sigma_u^-)$, and $O^*$ corresponding to variations in γa or γb, respectively. From figure 2, with the increase in γa or γb, the densities of both $O_2(a^1\Delta g)$ and $O_2(b^1\Sigma g^+)$ decrease accordingly. Comparing these two cases, changes in γb have a greater impact on the density of $O_2(b^1\Sigma g^+)$. Specifically, as γb increases from its lowest to its highest value, its density decreases by approximately 15 times, whereas increasing γa leads to about a 9-fold decrease in the density of $O_2(a^1\Delta g)$. Moreover, when γa or γb are low, their densities can become the highest among all neutral species, further underscoring the significance of these two metastable states. When γa is adjusted, the densities of O, $O_2(b^1\Sigma g^+)$, and $O_2(A^3\Sigma_u^+, A^3\Delta_u, c^1\Sigma_u^-)$ remain almost unchanged, the $O^*$ density increases slightly, and the $O_3$ density decreases. However, when γb increases, the densities of O, $O_2(a^1\Delta g)$, $O_2(A^3\Sigma_u^+, A^3\Delta_u, c^1\Sigma_u^-)$, and $O^*$ increase, by approximately 30%-40%, while the $O_3$ density increases by almost two orders of magnitude from $2.4\times10^{15}$ m$^{-3}$ to $1.1\times10^{17}$ m$^{-3}$. Therefore, compared to changes in γa, varying γb has a more significant impact on the densities of neutral species.

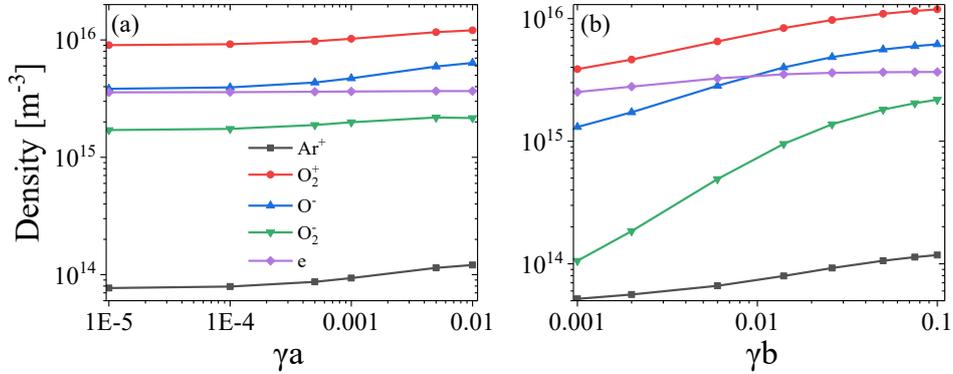

Figure 5 $Ar^+$ density (black line), $O_2^+$ density (red line), $O^-$ density (blue line), $O_2^-$ density (green line), electron density (purple line), as a function of γa (a), and γb (b). The discharge conditions are the same as those in figure 2.

Figures 4 and 5 show the variations in electronegativity and charged particle densities as γa or γb changes, respectively. From figure 5, the electronegativity increases with the increasing γa or γb, and changes in γb have a more pronounced effect on the electronegativity. From figure 5, increasing γa or γb results in a general increase in the densities of charged particles, and the variations in charged particle densities are more pronounced than those of neutral species. This is attributed to the significant increase in electronegativity, which enhances electron heating and the electron temperature, thereby intensifying ionization and other electron dynamics-related reactions. However, a comparison between figures 5 (a) and (b) also reveals that changes in γb have a more significant impact on the densities of charged particles. Moreover, the electron density remains nearly unchanged with the variation of γa and γb. For example, as γb increases, the electron density increases only from $2.51\times10^{15}$ m$^{-3}$ to $3.65\times10^{15}$ m$^{-3}$ by approximately 40%.

The above discussion highlights the significance of $O_2(a^1\Delta g)$ and $O_2(b^1\Sigma g^+)$, especially the often-overlooked $O_2(b^1\Sigma g^+)$. Therefore, this provides a valuable reference for future simulation studies and underscores the necessity of including reactions involving $O_2(b^1\Sigma g^+)$ in the reaction sets for $O_2$ or $O_2$-containing gas mixtures.

**3.2 Effect of γa or γb depending on different materials on Ar/O$_2$ discharge characteristics**

In this section, the influence of γa or γb depending on different materials and pressures on discharge characteristics is investigated. Figures 6 and 7 present the electron temperature, electronegativity and the densities of $O_2(a^1\Delta g)$ as well as $O_2(b^1\Sigma g^+)$ as a function of different material and pressure. In figure 6, the electron temperature gradually decreases with increasing pressure primarily due to the higher frequency of collisions between electrons and neutral particles at higher pressures. And, as the pressure increases, the electronegativity declines, except in the case of Teflon related quenching coefficients at the pressure range from 20 mTorr to 40 mTorr. The reason for the change in electronegativity will be explained in detail subsequently. At the same pressure, the electron temperature based on the stainless steel related quenching coefficients is higher than in the other two cases. Moreover, the electronegativity is lowest when the quenching coefficient of stainless steel is used, followed by quenching coefficient of copper, and highest in the case of that of Teflon.

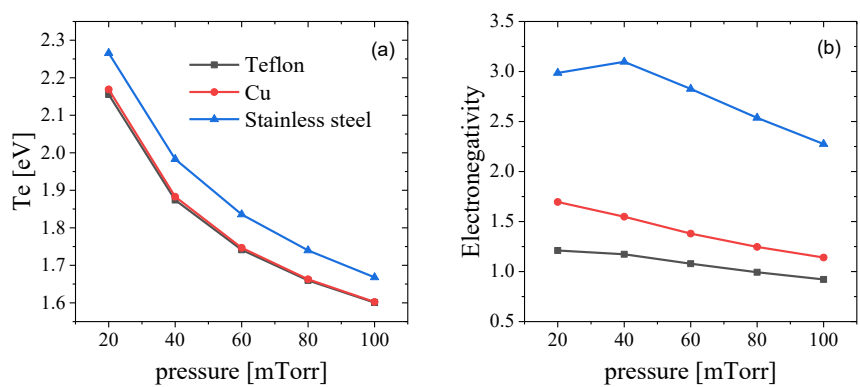

Figure 6 Electron temperature (a) and electronegativity (b) as a function of the pressure and quenching coefficients based on Teflon (black), Cu (red), and Stainless steel (blue). Discharge condition: gas ratio of Ar/O$_2$ is 50/50, with an electrode gap of 4.5 cm, a pressure range of 20-100 mTorr, a frequency of 13.56 MHz, and a voltage amplitude of 150 V.

From figure 7, for different quenching coefficient, the densities of both $O_2(a^1\Delta g)$ and $O_2(b^1\Sigma g^+)$ show an increasing trend as the pressure increases. Under the same pressure, the densities of $O_2(a^1\Delta g)$ and $O_2(b^1\Sigma g^+)$ are the lowest when the quenching coefficient is based on stainless steel. This is mainly because there are larger quenching coefficients associated with stainless steel. The lower densities of $O_2(a^1\Delta g)$ and $O_2(b^1\Sigma g^+)$ further attenuate the detachment reactions between the $O_2(a^1\Delta g)$ or $O_2(b^1\Sigma g^+)$, and $O^-$ ions, leading to a higher electronegativity and, consequently, a higher electron temperature (as shown in figure 6). When the quenching coefficients depending on copper or Teflon, the electronegativity shows little difference between the two scenarios, however, it is slightly higher in the case based on copper. The density of $O_2(a^1\Delta g)$ is relatively high in the case related to copper, but its corresponding quenching coefficient on the copper surface is lower. Meanwhile, the density of $O_2(b^1\Sigma g^+)$ is relatively low. These factors together lead to a reduced loss of negative ions and a slightly higher electronegativity.

Additionally, among these three sets of quenching coefficients, copper-related $\gamma a$ is lowest, resulting in the highest density of $O_2(a^1\Delta g)$. Similarly, Teflon has the lowest $\gamma b$, leading to the highest density of $O_2(b^1\Sigma g^+)$. It can be also observed that when the material is Teflon, the densities of $O_2(a^1\Delta g)$ and $O_2(b^1\Sigma g^+)$ are comparable. For example, at 20 mTorr, the density of $O_2(a^1\Delta g)$ is $1.03\times10^{19}$ m$^{-3}$, while that of $O_2(b^1\Sigma g^+)$ is $9.26\times10^{18}$ m$^{-3}$.

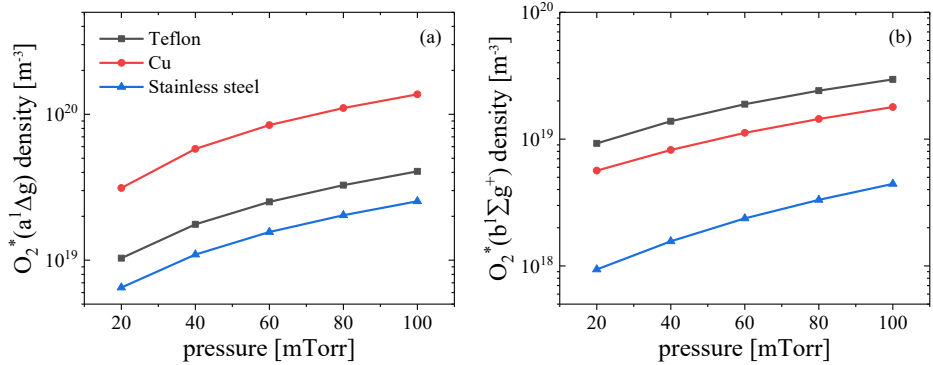

Figure 7 Densities of $O_2(a^1\Delta g)$ (a), $O_2(b^1\Sigma g^+)$ (b) as a function of the pressure and quenching coefficients depending on the different materials. The discharge conditions are the same as those in figure 7.

Figure 8 shows the densities of electrons, $O^-$, $O_2^-$, $Ar^+$, and $O_2^+$ as a function of the quenching coefficients and pressure. A comparison for quenching coefficients reveals that the densities of the major charged species are highest when the quenching coefficients is based on stainless steel and lowest when it is related to Teflon. As the pressure increases, the trend of the electron density shows a decrease followed by an increase in the case of stainless steel quenching coefficient and copper quenching coefficient of $O_2(a^1\Delta g)$, $O_2(b^1\Sigma g^+)$. In contrast, the case of Teflon-related quenching coefficient exhibits the different behavior, demonstrating an increasing trend as the pressure increases. In addition, the trend of the O⁻ density is consistent with a gradual decrease for different set of quenching coefficients. From figure 8 (c), the $O_2^-$ density shows different trends with increasing pressure for the three sets of quenching coefficients. For the case related to copper, the $O_2^-$ density gradually decreases. In the case depending on Teflon, the $O_2^-$ density remains nearly constant across the pressure range. In contrast, for the case related with stainless steel, the $O_2^-$ density initially increases, reaches a peak at 40 mTorr, and then gradually decreases. Additionally, by comparing figures 9 (a), (b), and (c), it can be concluded that, compared to the electron density and O⁻ density, the $O_2^-$ density consistently plays a secondary role.

For example, in the case of the copper, O⁻ is the dominant negative ion in the Ar/O₂ discharge at low pressures, while at higher pressures (≥80 mTorr), the electron density surpasses the O⁻ density. For the stainless steel cases, the O⁻ density remains significantly higher than both the electron and $O_2^-$ densities across the entire pressure range. However, with the Teflon cases, the electron and O⁻ densities are relatively comparable under different pressures.

In figure 8 (d), it can be seen that as the pressure increases, the $Ar^+$ density decreases with increasing pressure. From figure 8 (e), as the pressure increases, the $O_2^+$ density shows different trends depending on the $\gamma_a$ or $\gamma_b$. With a copper material, the $O_2^+$ density remains nearly constant. For Teflon cases, it initially increases and then becomes nearly unchanged. However, for stainless steel, the $O_2^+$ density initially increases and then decreases as the pressure increases.

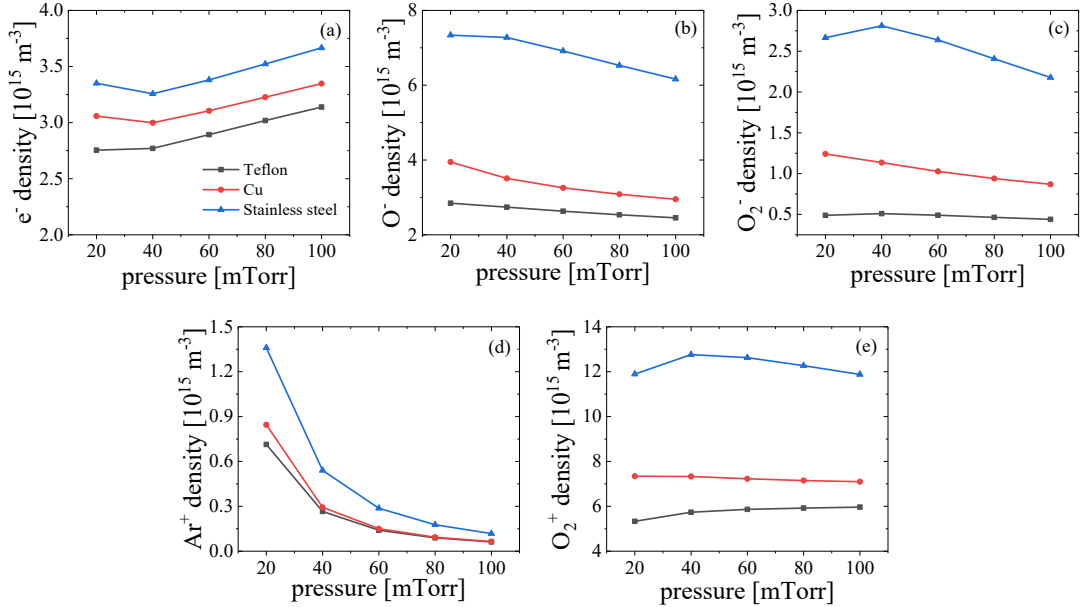

Figure 8 Densities of electron (a), $O_2^-$ (b), $O_2^-$ (c), $Ar^+$ (d), $O_2^+$ (e) as a function of the pressure and quenching coefficients depending on the different materials. The discharge conditions are the same as those in figure 7

From figure 6 and figure 7, for the Teflon cases, the electronegativity remains nearly constant with increasing pressure at a pressure of less than 40 mTorr, primarily because both the electron density and O⁻ density remain unchanged. In the case of copper, although the electron density (figure 8 (a)) decreases slightly, the significant reduction in both O⁻ and $O_2^-$ densities lead to a decrease in electronegativity as the pressure increases. For the case of stainless steel material, as the pressure increases from 20 mTorr to 40 mTorr, the electron density decreases slightly, the $O_2^-$ density shows an increasing trend, and the O⁻ density remains nearly constant, which results in a slight increase in electronegativity. And when the pressure further increases, the electron density enhances, while the densities of O⁻ and $O_2^-$ decline, which causes a reduction in electronegativity for these three cases made of different materials.

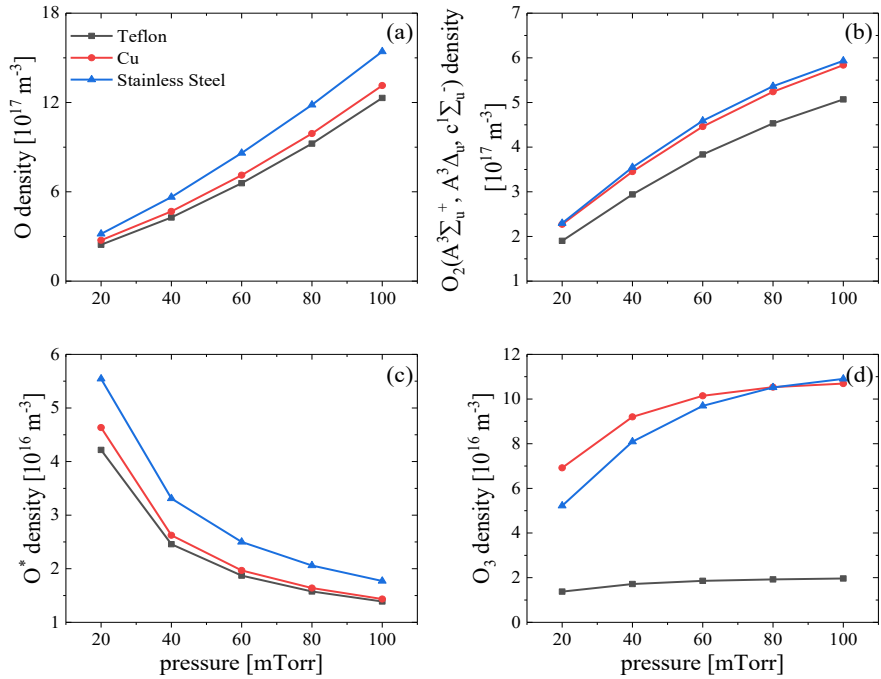

Figure 9 Densities of O (a), $O_2(A^3\Sigma_u^+, A^3\Delta_u, c^1\Sigma_u^-)$ (b), $O^*$ (c), and $O_3$ (d) as a function of the pressure and different materials. The discharge conditions are the same as those in figure 7

The densities of O, $O_2(A^3\Sigma_u^+, A^3\Delta_u, c^1\Sigma_u^-)$, $O^*$, and $O_3$ for different γa, γb depending on the materials and pressures are presented in figure 9. For different γa, γb depending on different materials, the variation trends of the neutral densities with increasing pressure are generally consistent. Specifically, as the pressure increases, the densities of O, $O_2(A^3\Sigma_u^+, A^3\Delta_u, c^1\Sigma_u^-)$, and $O_3$ gradually increase, while the density of $O^*$ decreases. Furthermore, when the cases are based on Teflon or copper, the densities of O and $O^*$ are comparable at lower pressure (< 40 mTorr), and both particle densities are lower than those observed for a stainless steel material in the whole pressure range. In contrast, in the cases depending on stainless steel or copper materials, the magnitude of the densities of $O_2(A^3\Sigma_u^+, A^3\Delta_u, c^1\Sigma_u^-)$ and $O_3$ are significantly higher than those in the case of Teflon. It can also be observed that when the case is based on copper, the $O_3$ density is slightly higher than that in the presence of case depending on stainless steel material at lower pressures (< 80 mTorr). As the pressure increases further, the difference in $O_3$ density between the two cases becomes minimal.

The changes in particle density can be further explained and explored based on particle generation and loss rates. Since the change of particle densities as a function of pressure had been discussed in detail based on the stainless steel material in previous work according to their generation and loss rates [37], this study will mainly focus on the variation of the individual particle generation and loss terms for different γa and γb depending on different materials. Figure 10 shows the variation of the generation and loss rates of electrons, $O^-$, $O_2^-$ for different γa and γb, in which, reactions that contribute less than 5% to the electron, $O^-$, and $O_2^-$ generation or loss are neglected. Besides, compared to a case based on stainless steel material, the trends of the densities of the three negatively charged particles differ slightly in the cases of Teflon and copper material at lower pressures. Therefore, results at 20 mTorr and 40 mTorr are provided. The relevant reactions are given in the Appendix table A1 and A2. In figures 11 (a1) and (a2), the loss rate of electrons to the wall, k, is numerically equal to the sum of the loss rate for each positive ion to the wall, i.e., the sum of the loss rate for reactions R100, R101, and R102.

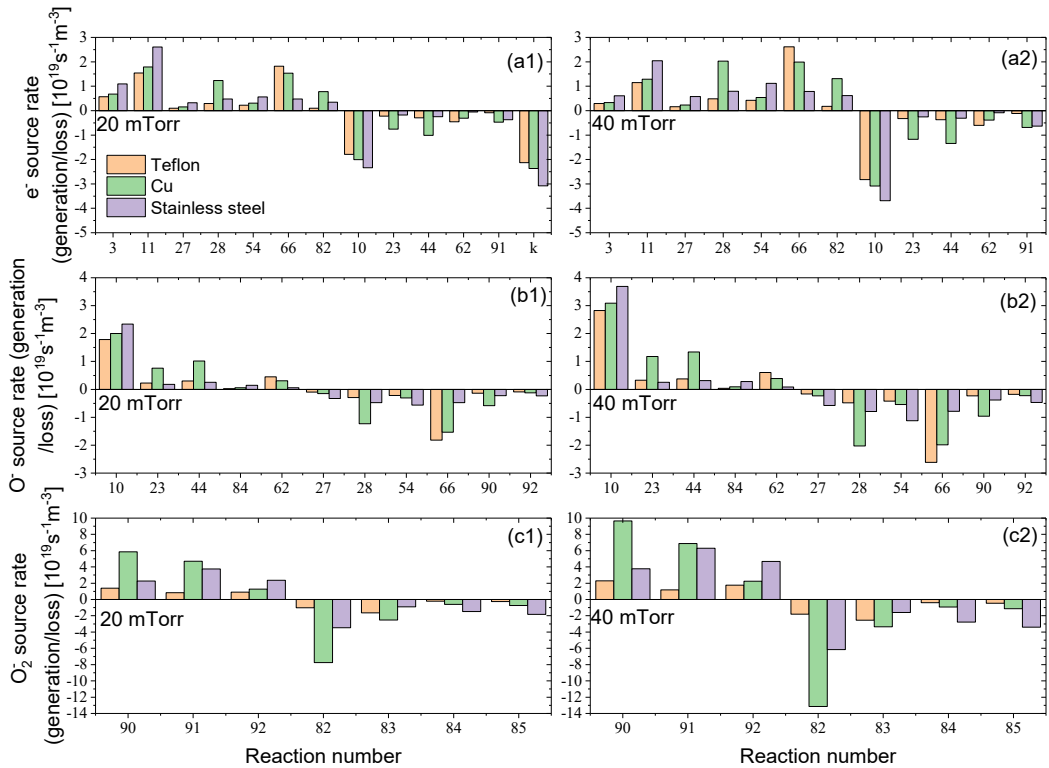

Figure 10 Generation and loss rate of electrons (a1), (a2), $O^-$ (b1), (b2), and $O_2^-$ (c1), (c2) for different different materials at 20 mTorr, and 40 mTorr, k is the electron loss rate to the wall. Note: The values of the horizontal coordinates correspond to those labeled in the reaction list

Comparing two scenarios based on Teflon and copper material, due to the higher density of $O_2(a^1\Delta g)$ for the copper case, the electron generation or loss rates of reactions R23, R28, R44, and R82 involving $O_2(a^1\Delta g)$ are significantly higher. Besides, since the $O_3$ density is also higher in the case of copper material, as shown in figure 9 (d), the reaction rate of R91 (e+$O_3 \rightarrow$ O+$O_2^-$) is correspondingly higher than that in the case based on Teflon material. In addition, for electron generation, except for a slight decrease in the reaction rate of R66 ($O^-$+$O_2(b^1\Sigma g^+) \rightarrow O_2$+O+$e^-$), the results show that the generation rates are higher in the cases depending on copper material. This leads to a lower electron density in the case based on Teflon material compared to that in the copper material. Compared with the copper material cases, in the case based on stainless steel material, the densities of $O_2(a^1\Delta g)$ and $O_2(b^1\Sigma g^+)$ are lower. As a result, the rates of reactions R28, R23, R44, R66, and R82, which involve these two species, are correspondingly lower. Meanwhile, with the cases based on stainless steel material, the electron temperature is slightly higher, leading to an enhanced electron-impact ionization (R11) rate between electrons and background $O_2$. Moreover, due to the higher $O^-$ density in the stainless steel material cases, the reaction rates of electron generation reactions R27 and R54 are accordingly higher. In summary, compared with the case depending on copper material, although the rates of electron-generation reactions R28, R82, and R66 are lower, the higher rates of R11, R27, and R54, combined with the lower rates of electron loss reactions (R23 and R44), result in a higher electron density in the case based on stainless steel material. In contrast, for Teflon material cases, the electron density remains nearly constant in the pressure range of 20–40 mTorr, compared with the other two materials cases. As the pressure increases, in the case based on Teflon, the rates of electron-impact ionization reactions (R3 and R11) decrease, while the rate of the dissociative attachment reaction R66 increases. The opposing effects on electron density offset each other, resulting in a constant electron density within this pressure range. However, in the cases depending on copper and stainless steel, the rates for associative detachment of $O^-$ by $O_2(b^1\Sigma g^+)$ and O (R27, R28), as well as detachment reactions (R54, R66, R82), increase, but the rates for electron impact ionization (R3, R11) decline, and attachment rate (R10) increases at 40 mTorr, further resulting in a lower electron density compared to the results observed at 20 mTorr. Moreover, at 100 mTorr, similar to previous work [37], based on these three materials, associative detachment and detachment reactions which can generate electrons become the predominant mechanisms for electron generation, leading to higher electron density.

In figure 10 (b1) and (b2), compared with the case depending on Teflon material, the use of copper related coefficient results in a higher density of $O_2(a^1\Delta g)$, which leads to higher rates for R23, R28, R44, and R90 involving $O_2(a^1\Delta g)$. Although this enhances the rates of $O^-$ detachment loss reactions (R28 and R90), the effect of generation reactions of $O^-$ (R10, R23 and R44) is more pronounced, ultimately yielding a higher $O^-$ density in the cases depending on the copper. In comparison to the case of copper, the scenario of stainless steel as the different material exhibits significantly lower densities of both $O_2(a^1\Delta g)$ and $O_2(b^1\Sigma g^+)$, resulting in reduced reaction rates for R23, R44, R28, R62, R66, and R90. However, in the case based on stainless steel, the electron-impact dissociation reaction R10 involving $O_2$ is stronger, and loss rates (R28, R66, and R90) are lower. These reasons lead to a higher $O^-$ density for the case depending on stainless steel material. In contrast, in the cases based on copper, the $O^-$ density decreases significantly as the pressure rises from 20 to 40 mTorr. This is due to the significant increase in the reaction rates of $O^-$ loss reactions R28, R66, and R90 as the pressure increases from 20 mTorr to 40 mTorr and even to higher pressure (100 mTorr). For the cases depending on stainless steel and Teflon, the change of the $O^-$ density is dominated by an increase in the $O^-$ loss reaction rate when the pressure increases from 20 mTorr to 100 mTorr.

From figure 10 (c1) and (c2), comparing the case depending on copper to that of Teflon, although the loss reaction rate of $O_2^-$ (R82) is relatively high, its generation reaction rates (R90 and R91) are significantly high and play a dominant role, causing the $O_2^-$ density to be higher with copper different than that in the case of Teflon. When stainless steel is considered in the case instead of copper, the densities of $O_2(a^1\Delta g)$ and $O_2(b^1\Sigma g^+)$ are notably lower. This causes that the generation rate of $O_2^-$ (R90) is reduced, the more substantial decrease in the loss reaction rates R82 and R83 can be observed, resulting in a higher $O_2^-$ density for the case of stainless steel. As the pressure increases from 20 mTorr to 40 mTorr, the trends in density of $O_2^-$ differs among three scenarios. For Teflon scenario, both the generation reactions (R90 and R91) and the loss reactions (R82 and R83) of $O_2^-$ increase slightly with the increasing pressure. These opposing effects offset, keeping the $O_2^-$ density nearly constant. In the copper scenario, the loss rate of $O_2^-$ (R82) increases significantly as pressure enhances, leading to a reduction in $O_2^-$ density. For stainless steel scenario, as pressure increase to 40 mTorr the $O_2^-$ generation rate (R90, R91, R92) increases significantly compared to its loss rate, which results in higher $O_2^-$ density. However, the loss rate increases significantly, leading to a decreasing trend of $O_2^-$ density, as the pressure further increases.

Figure 11 presents the variation in main generation and loss rates of $O_2^+$ at 20 mTorr, 40 mTorr, and 100 mTorr for different γa and γb, in which, reactions that contribute less than 5% to the $O_2^+$ generation or loss are ignored. Comparing the three different materials, it is evident that as the scenarios changes from Teflon to copper or stainless steel, the generation rate of $O_2^+$ (R11) gradually enhances, resulting in a corresponding higher $O_2^+$ density. This is because the proportion of high-energy electron or electron temperature is higher in the stainless steel scenarios, and high-energy electron are needed in the generation reaction of $O_2^+$. As pressure increases, the $O_2^+$ densities under these three scenarios exhibit different trends. As pressure increases, the generation and loss rates of $O_2^+$ all declines. For Teflon scenario, increasing pressure leads to reduced wall loss coefficient (R101). This may be related to the decrease in Bohm velocity caused by the decrease in electron temperature. And the loss rate of $O_2^+$ decreases more rapidly. Therefore, the $O_2^+$ density needs to increase to maintain enough loss of $O_2^+$ density to balance between generation and loss of $O_2^+$. In contrast, for copper scenario, the $O_2^+$ density remains relatively unchanged due to the simultaneous decrease in both generation and loss reaction rates. And the variation of $O_2^+$ density in stainless steel scenario has been analyzed in previous work [37] and will not be repeated here.

For $Ar^+$ ions, under all three scenarios depending on different material, the ion density declines as pressure increases. The generation of $Ar^+$ typically requires high-energy electrons. However, higher pressure enhances collisions between electrons and neutral particles, reducing the proportion of high-energy electrons. Therefore, $Ar^+$ density decreases as the pressure increases. Compared to the other scenarios, in the case depending on stainless steel, the electron temperature, or the proportion of high-energy electrons is relatively higher. This leads to a higher reaction rate of the $Ar^+$ generation rate (R3), resulting in a higher $Ar^+$ density.

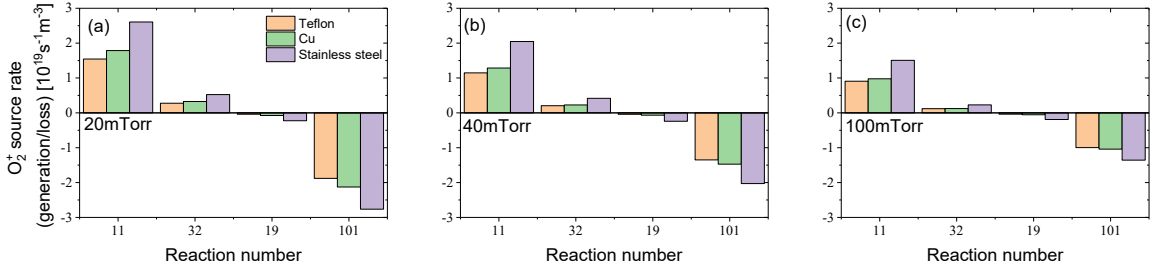

Figure 11 Generation and loss rate of $O_2^+$ under different quenching coefficients and pressure. Note: The values of the horizontal coordinates correspond to those labeled in the reaction list.

The neutral species densities exhibit similar trends with increasing pressure across different γa and γb, so the related discussion can refer to our previous work [37]. Therefore, only the reaction rates of various neutral species at 20 mTorr are analyzed to explain the influence of γa and γb on their densities. Figure 12 presents the generation and loss rates of O, $O_2(A^3\Sigma_u^+, A^3\Delta_u, c^1\Sigma_u^-)$, $O^*$, and $O_3$ under different γa and γb, in which, reactions that contribute less than 5% to the O, $O_2(A^3\Sigma_u^+, A^3\Delta_u, c^1\Sigma_u^-)$, $O^*$, and $O_3$ generation or loss are overlooked. For O and $O^*$, their densities are all highest in stainless steel scenario, next highest in copper scenario, and lowest under Teflon scenario, as shown in figure 9. From figures 13 (a) and (c), this is because their main generation rates are all higher in the stainless steel scenario and copper scenario, compared with Teflon cases.

From figure 12 (b), when comparing the results in copper scenario to Teflon scenario, the generation rate of R73 is high under stainless steel scenario, but the significantly higher $O_2(a^1\Delta g)$ density under copper cases leads to a higher rate of R75. Therefore, the above reasons lead to a small difference in the density of $O_2(A^3\Sigma_u^+, A^3\Delta_u, c^1\Sigma_u^-)$ under the two scenarios. In the case of Teflon, the generation rate of $O_2(A^3\Sigma_u^+, A^3\Delta_u, c^1\Sigma_u^-)$ is the lowest, and therefore its density is lower compared to the other two scenarios. From figure 3, since the highest $O_2(a^1\Delta g)$ density can be observed in copper scenario, and this species is involved in the generation of $O_3$ through reaction R28, the rate of R28 is significantly higher under copper scenario (figure 12 (d)), further leading to the highest $O_3$ density. However, as the pressure increases, the difference in $O_3$ between the case of copper and stainless steel declines. From figures 2 and 4, with the increasing pressure, $O^-$ density decreases, $O_2(a^1\Delta g)$ density increases, and the two factors reduces the difference related to R28. When comparing stainless steel case to Teflon case, the higher densities of $O^-$, O, and $O_2^-$ under stainless steel case enhance the reaction rates of $O_3$ generation reactions R28 and R85. As a result, the $O_3$ density is higher in the stainless steel scenario compared to the case of Teflon.

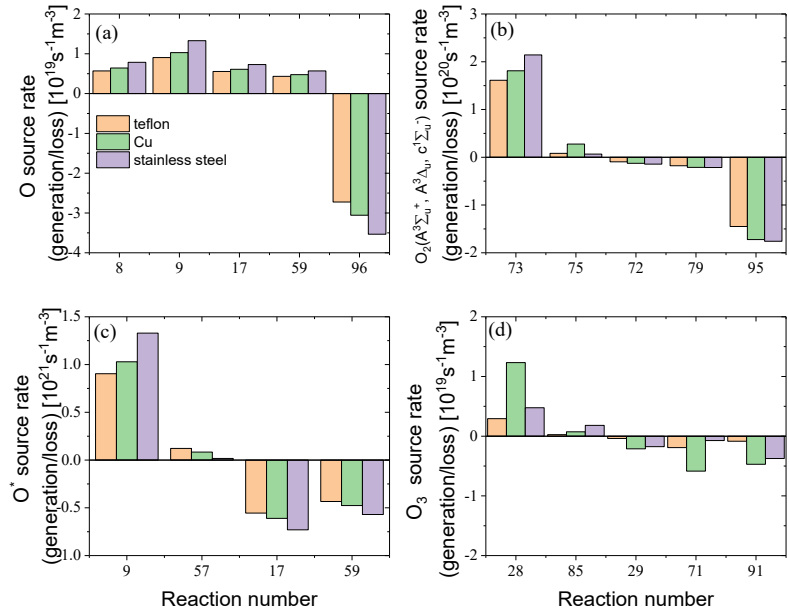

Figure 12 Generation and loss rate of O (a), $O_2(A^3\Sigma_u^+, A^3\Delta_u, c^1\Sigma_u^-)$ (b), $O^*$ (c), and $O_3$ (d) as a function of quenching coefficients at 20 mTorr. Note: The values of the horizontal coordinates correspond to those labeled in the reaction list.

The inductance and resistance of the plasma bulk region is a key physical parameter characterizing the plasma's electrical conductivity. For example, changes in this resistance directly influence energy dissipation within the plasma, thereby affecting the discharge efficiency. Figure 13 shows the inductance and resistance of the plasma bulk region, as well as the sheath voltage drop, under different γa, γb and pressures. Under the same pressure, a comparison of the results indicates that in the case of stainless steel, the bulk inductance, resistance, and sheath voltage drop are all at their lowest values. In contrast, these values are highest in the Teflon case. The inductance and resistance of the plasma bulk region are inversely related to the electron density [37]. As shown in figure 8, the electron density is highest with stainless steel scenario, followed by copper scenario, and lowest with Teflon scenario, which directly explains the observed trends in inductance and resistance. Furthermore, since the sheath voltage drop is inversely proportional to the positive ion density, the highest positive ion density observed in stainless steel scenario, as shown in figure 8, leads to the highest corresponding sheath voltage drop. In addition, Figure 13 shows that for different quenching coefficient depending on material properties, the inductance and resistance of the plasma bulk region, as well as the sheath voltage drop, exhibit consistent trends with increasing pressure. The variation in inductance with pressure is also inversely proportional to the change in electron density. Moreover, as the pressure increases, the effective collision frequency between electrons and neutral particles enhances, resulting in an increase in resistance. Additionally, according to Ref. [37], the reduction of charge within the sheath becomes more pronounced at higher pressures, leading to a further decrease in the sheath voltage drop.

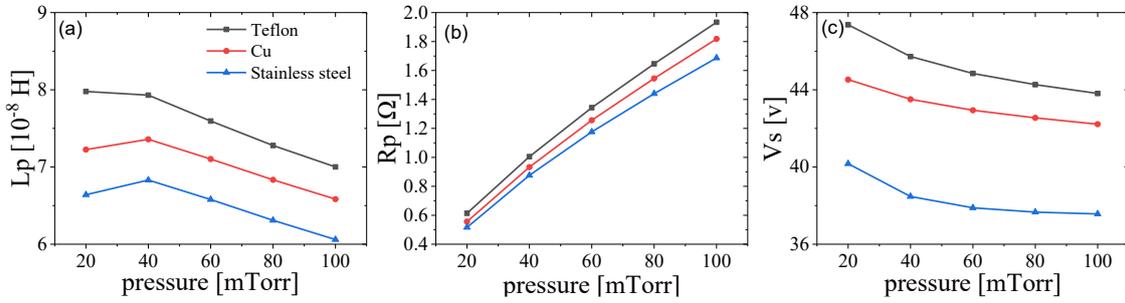

Figure 13 The inductance (a) and the resistance (b) of the plasma bulk, and the voltage drop across the sheath (c) as a function of the pressure and quenching coefficients depending on materials. The discharge conditions are the same as those in figure 7

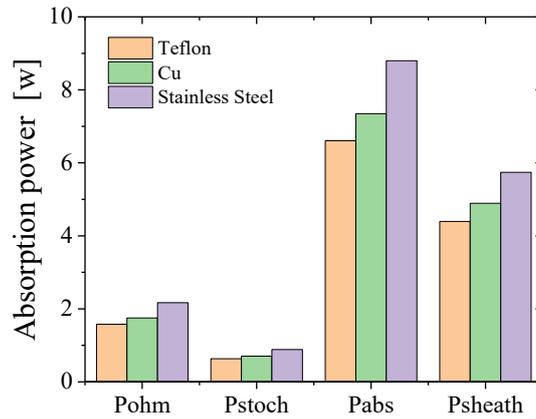

Figure 14 Ohmic heating $P_{ohm}$, stochastic heating $P_{stoch}$, sheath dissipation $P_{sheath}$, and total absorption power in plasma $P_{abs}$ under different quenching coefficients depending on materials at 20 mTorr. The discharge conditions are the same as those in figure 7

Based on the heating mechanism, the total absorbed power in plasma $P_{abs}$ can be divided into three parts, which are the ohmic heating, stochastic heating and sheath dissipation. Figure 14 presents the Ohmic heating power $P_{ohm}$, stochastic heating power $P_{stoch}$, sheath power dissipation $P_{sheath}$, and the total absorbed power in the plasma $P_{abs}$ at a pressure of 20 mTorr for different $\gamma_a$, $\gamma_b$ depending on different materials. From figure 14, $P_{sheath}$ contributes the most to the total power, while $P_{ohm}$ and $P_{stoch}$ play minor roles. Besides, the power losses in various regions of the plasma, as well as the total absorbed power, are highest in the case of stainless steel and lowest in the scenario of Teflon. $P_{sheath}$ is mainly the energy gained by positive ions as they are accelerated by the voltage drop across the sheath. From figure 13 (c), voltage drop across the sheath for different quenching coefficients depending on materials show the opposite trends. But, $P_{sheath}$ is also related to the radio frequency current $I_{rf}(t)$ and the phase difference between voltage drop across the sheath $V_s(t)$ and $I_{rf}(t)$. At 20 mTorr, the peak of plasma current during one RF cycle is 3.45 A for stainless steel, 2.99 A for copper, and 2.70 A for Teflon. This trend of plasma current can lead to higher $P_{sheath}$ in the case of stainless steel, despite the lower voltage drop across the sheath in this case. $P_{ohm}$ and $P_{stoch}$ are also related to and affected by the plasma current. This means the trend of plasma current explains the change in total absorbed power and its components with different scenarios. As pressure increases, there are a noticeable increase in $P_{ohm}$, and a decrease in $P_{sheath}$, due to the intense collisions between electrons and neutrals as well as the declined voltage drop across the sheath. Meanwhile, there is almost no change in the $P_{stoch}$. Besides, $P_{ohm}$ is one of the most essential components contributing to $P_{abs}$. These factors combine to result in a higher $P_{abs}$. The detailed analysis of the evolution of power absorption and its components with varying pressure can be found in our previous work [37].

## 4 Conclusions

In this work, the effects of variations the quenching coefficients of $O_2(a^1\Delta g)$ and $O_2(b^1\Sigma g^+)$ on the characteristics of capacitively coupled $Ar/O_2$ discharges are systematically investigated. As the quenching coefficients of $O_2(a^1\Delta g)$ or $O_2(b^1\Sigma g^+)$ increase, the densities of these metastable species decrease. This weakens the detachment reactions between the $O_2(a^1\Delta g)$, $O_2(b^1\Sigma g^+)$ and $O^-$, leading to an increase in both the electron temperature and the plasma electronegativity. This further cause that the densities of $Ar^+$, $O_2^+$, and $O^-$ increase significantly. For neutral species, varying $\gamma_a$ has minimal impact on the densities of O, $O_2(b^1\Sigma g^+)$, $O_2(A^3\Sigma u^+, A^3\Delta u, c^1\Sigma u^-)$, and $O^*$. In contrast, varying $\gamma_b$ results in an increase in the densities of O, $O_2(a^1\Delta g)$, $O_2(A^3\Sigma u^+, A^3\Delta u, c^1\Sigma u^-)$, and $O^*$, while the $O_3$ density increases by approximately two orders of magnitude. Increasing the $\gamma_b$ also leads to a moderate increase in electron density, whereas changing the $\gamma_a$ has almost no effect on electron density. These results highlight that $O_2(b^1\Sigma g^+)$ also plays a significant role in $Ar/O_2$ discharges and should not be neglected.

This work also investigates the effects of quenching coefficients and pressures on particle densities and discharge characteristics, with a particular focus on the quenching coefficients of $O_2(a^1\Delta g)$ and $O_2(b^1\Sigma g^+)$ according to different materials. The study reveals that among the three materials, stainless steel scenario results in the lowest densities of both $O_2(a^1\Delta g)$ and $O_2(b^1\Sigma g^+)$. This reduction weakens the detachment reactions between the $O_2(a^1\Delta g)$ (or $O_2(b^1\Sigma g^+)$) and $O^-$, leading to enhanced electronegativity and, consequently, a higher electron temperature. In contrast, copper scenario yields the highest density of $O_2(a^1\Delta g)$, while Teflon scenario leads to the highest density of $O_2(b^1\Sigma g^+)$. Moreover, in the Teflon scenario, the densities of $O_2(a^1\Delta g)$ and $O_2(b^1\Sigma g^+)$ are nearly equal and are the highest among all neutral species, highlighting their dominant role under Teflon scenarios. The difference in quenching coefficients of $O_2(a^1\Delta g)$ and $O_2(b^1\Sigma g^+)$ according to different materials will further affect the density of individual charged and neutral particles. In this work, the variation of the density of individual particles under different quenching coefficients of $O_2(a^1\Delta g)$ and $O_2(b^1\Sigma g^+)$ is extensively analyzed. In general, the variation of these particle densities is correlated with the effect of the quenching coefficients, the $O_2(a^1\Delta g)$ and $O_2(b^1\Sigma g^+)$ densities, as well as the electronegativity, and the electron temperature. The evolution of the individual particle densities is systematically explained by the variation of the above parameters combined with the particle generation and loss channels.

Moreover, the inductance, resistance, and voltage drop across the sheath are analyzed for different quenching coefficients depending on materials. In the case of stainless steel, the inductance, resistance, and sheath voltage drop are lowest, while they are highest when the quenching coefficients of $O_2(a^1\Delta g)$ and $O_2(b^1\Sigma g^+)$ is based on Teflon. Additionally, the power losses in various parts of the plasma, as well as the total absorbed power, are highest in stainless steel cases and lowest in Teflon cases. This is mainly because the plasma current is higher under stainless steel cases. In the future, we will carry out the effect of tailored voltage waveform on discharge characteristics and particle generation or loss based on this model.

**Acknowledgments**

This work was supported by the National Natural Science Foundation of China under Grant Nos. 12475202, 12405289, the China Scholarship Council (No. 202306060179).

**Appendix**

Table A1 Collision reactions, reaction rate coefficients, and references considered in this work.

| # | Reactions | Rate coefficients ($m^3s^{-1}$) | Reference |
|---|---|---|---|
| 1 | e + Ar → Ar + e | a | [43] |
| 2 | e + Ar → Ar$^*$ + e | $9.73\times10^{-16}\times T_e^{-0.07}\times\exp(-11.69/T_e)$ | [43] |
| 3 | e + Ar → Ar$^+$ + 2e | $2.39\times10^{-14}\times T_e^{-0.57}\times\exp(-17.43/T_e)$ | [43] |
| 4 | e + Ar$^*$ → Ar$^+$ + 2e | $2.05\times10^{-13}\times\exp(-4.95/T_e)$ | [43] |
| 5 | e + Ar$^*$ → Ar + e | $2.0\times10^{-13}$ | [43] |
| 6 | Ar$^*$ + Ar$^*$ → Ar+Ar$^+$ + e | $6.2\times10^{-16}$ | [43] |

| 7 | $e + O_2 \rightarrow O_2 + e$ | $\exp(-30.7683+0.3531\times\ln(T_e)+0.2068\times(\ln(T_e))^2-0.0406\times(\ln(T_e))^3)$ | [43] |
| --- | --- | --- | --- |
| 8 | $e + O_2 \rightarrow O + O + e$ | $4.23\times10^{-15}\times\exp(-5.56/T_e)$ | [43] |
| 9 | $e + O_2 \rightarrow O + O^* + e$ | $5.0\times10^{-14}\times\exp(-8.4/T_e)$ | [43] |
| 10 | $e + O_2 \rightarrow O + O^-$ | $1.07\times10^{-15}\times T_e^{-1.39}\times\exp(-6.26/T_e)$ | [43] |
| 11 | $e + O_2 \rightarrow O_2^+ + 2e$ | $2.34\times10^{-15}\times T_e^{-1.03}\times\exp(-12.9/T_e)$ | [43] |
| 12 | $e + O \rightarrow O + e$ | $\exp(-30.9463+0.9484\times\ln(T_e))-0.14158(\ln(T_e))^2-0.0154\times(\ln(T_e))^3$ | [43] |
| 13 | $e + O \rightarrow O^* + e$ | $4.47\times10^{-15}\times\exp(-2.286/T_e)$ | [43] |
| 14 | $e + O \rightarrow O^+ + 2e$ | $9.0\times10^{-15}\times T_e^{0.7}\times\exp(-13.6/T_e)$ | [43] |
| 15 | $e + O^* \rightarrow O^+ + 2e$ | $9.0\times10^{-15}\times T_e^{0.7}\times\exp(-11.6/T_e)$ | [43] |
| 16 | $e + O^- \rightarrow O + 2e$ | $1.73\times10^{-13}\times\exp(-5.76/T_e+7.3/T_e^2-3.48/T_e^3)$ | [43] |
| 17 | $O_2 + O^* \rightarrow O_2 + O$ | $4.1\times10^{-17}$ | [43] |
| 18 | $O^* + O \rightarrow O + O$ | $8.1\times10^{-18}$ | [43] |
| 19 | $O_2^+ + O^- \rightarrow O_2 + O$ | $1.5\times10^{-13}\times(300/T_g)^{0.5}$ | [43] |
| 20 | $O^+ + O^- \rightarrow O + O$ | $2.7\times10^{-13}\times(300/T_g)^{0.5}$ | [43] |
| 21 | $e + O_2^+ \rightarrow O^* + O$ | $2.2\times10^{-14}\times(T_e)^{-0.5}$ | [43] |
| 22 | $e + O_2 \rightarrow O_2^* + e$ | $1.7\times10^{-15}\times\exp(-3.1/T_e)$ | [43] |
| 23 | $e + O_2^*(a) \rightarrow O + O^-$ | $2.28\times10^{-16}\times\exp(-2.29/T_e)$ | [43] |
| 24 | $e + O_2^*(a) \rightarrow O_2^+ + 2e$ | $9.0\times10^{-16}\times\exp(-11.6/T_e)$ | [43] |
| 25 | $e + O_2^*(a) \rightarrow 2O + e$ | $4.2\times10^{-15}\times\exp(-4.6/T_e)$ | [43] |
| 26 | $e + O_2^*(a) \rightarrow O_2 + e$ | $5.6\times10^{-15}\times\exp(-2.2/T_e)$ | [43] |
| 27 | $O^- + O \rightarrow O_2 + e$ | $1.4\times10^{-16}$ | [43] |
| 28 | $O^- + O_2^*(a) \rightarrow O_3 + e$ | $1.0\times10^{-16}$ | [43] |
| 29 | $e + O_3 \rightarrow O_2 + O + e$ | $1.4\times10^{-14}$ | [43] |

| # | Reaction | Rate | Ref |
|---|---|---|---|
| 30 | $e + O_3 \rightarrow O_2 + O^-$ | $2.12 \times 10^{-15} \times \exp(-0.93/T_e)$ | [43] |
| 31 | $O_2 + Ar^* \rightarrow O_2 + Ar$ | $1.12 \times 10^{-15}$ | [43] |
| 32 | $O_2 + Ar^+ \rightarrow O_2^+ + Ar$ | $1.2 \times 10^{-17}$ | [43] |
| 33 | $O_2 + Ar^* \rightarrow O + O + Ar$ | $5.8 \times 10^{-17}$ | [43] |
| 34 | $O + Ar^* \rightarrow O + Ar$ | $8.1 \times 10^{-18}$ | [43] |
| 35 | $O + Ar^+ \rightarrow O^+ + Ar$ | $1.21 \times 10^{-17}$ | [43] |
| 36 | $O^- + Ar^+ \rightarrow O + Ar$ | $2.7 \times 10^{-17}$ | [43] |
| 37 | $O_2^+ + O^- \rightarrow 3O$ | $2.6 \times 10^{-14} \times (300/T_g)^{0.44}$ | [29] |
| 38 | $e + O_2 \rightarrow O^+ + O^- + e$ | $7.1 \times 10^{-17} \times T_e^{0.5} \times \exp(-17.0/T_e)$ | [29] |
| 39 | $e + O_2 \rightarrow O^+ + O + 2e$ | $1.88 \times 10^{-16} \times T_e^{1.699} \times \exp(-16.81/T_e)$ | [29] |
| 40 | $O^+ + O_2 \rightarrow O_2^+ + O$ | $2.1 \times 10^{-17} \times (300/T_g)^{0.5}$ | [29] |
| 41 | $e + O^* \rightarrow O + e$ | $8.17 \times 10^{-17} \times \exp(-0.4/T_e)$ | [29] |
| 42 | $e + O_2 \rightarrow O^* + O^* + e$ | $1.95 \times 10^{-16} \times T_e^{0.22} \times \exp(-12.62/T_e)$ | [29] |
| 43 | $e + O_2^*(a) \rightarrow O^+ + O + 2e$ | $1.88 \times 10^{-16} \times T_e^{1.699} \times \exp(-15.83/T_e)$ | [29] |
| 44 | $e + O_2^*(a) \rightarrow O^* + O^-$ | $9.93 \times 10^{-15} \times T_e^{1.437} \times \exp(-7.44/T_e)$ | [29] |
| 45 | $O^* + O_2 \rightarrow O + O_2^*(a)$ | $1.0 \times 10^{-18}$ | [29] |
| 46 | $e + O_2^*(a) \rightarrow O + O^* + e$ | $1.29 \times 10^{-14} \times T_e^{0.22} \times \exp(-11.64/T_e)$ | [29] |
| 47 | $O^+ + O_3 \rightarrow O_2 + O_2 + e$ | $1.0 \times 10^{-16}$ | [29] |
| 48 | $O + O_3 \rightarrow O_2 + O_2$ | $1.81 \times 10^{-17} \times \exp(2300/T_g)$ | [29] |
| 49 | $O_2 + O_3 \rightarrow O_2 + O_2 + O$ | $7.26 \times 10^{-16} \times \exp(11400/T_g)$ | [29] |
| 50 | $O_2^*(a) + O_3 \rightarrow O_2 + O_2 + O$ | $6.01 \times 10^{-17} \times \exp(-2853/T_g)$ | [29] |
| 51 | $O_2^+ + O^- \rightarrow 3O$ | $2.6 \times 10^{-14} \times (300/T_g)^{0.44}$ | [29] |
| 52 | $e + O_2 \rightarrow O^+ + O^- + e$ | $7.1 \times 10^{-17} \times T_e^{0.5} \times \exp(-17.0/T_e)$ | [29] |
| 53 | $O^+ + O_2^*(a) \rightarrow O_2^+ + O$ | $2 \times 10^{-17}$ | [34] |
| 54 | $O^- + O_2 \rightarrow O_2 + O + e$ | $2.4 \times 10^{-18}$ | [34] |

| # | Reaction | Rate | Ref |
|---|---|---|---|
| 55 | $O_2^*(a) + Ar^+ \rightarrow O_2^+ + Ar$ | $1.2\times10^{-17}$ | [44] |
| 56 | $O^* + Ar^+ \rightarrow O^+ + Ar$ | $1.2\times10^{-17}$ | [44] |
| 57 | $e + O_2^*(b) \rightarrow O + O^* + e$ | $3.49\times10^{-14}\times\exp(-4.29/T_e)$ | [29] |
| 58 | $e + O_2 \rightarrow O_2^*(b) + e$ | $3.24\times10^{-16}\times\exp(-2.218/T_e)$ | [29] |
| 59 | $O^* + O_2 \rightarrow O + O_2^*(b)$ | $2.56\times10^{-17}\times\exp(67/T_g)$ | [34] |
| 60 | $e + O_2^*(b) \rightarrow O^+ + O + 2e$ | $1.88\times10^{-16}\times T_e^{1.699}\times\exp(-15.183/T_e)$ | [29] |
| 61 | $e + O_2^*(a) \rightarrow O_2^*(b) + e$ | $3.24\times10^{-16}\times\exp(-1.57/T_e)$ | [29] |
| 62 | $e + O_2^*(b) \rightarrow O + O^-$ | $4.19\times10^{-15}\times T_e^{-1.376}\times\exp(-4.54/T_e)$ | [29] |
| 63 | $e + O_2^*(b) \rightarrow O_2^+ + 2e$ | $2.34\times10^{-15}\times T_e^{1.03}\times\exp(-10.633/T_e)$ | [29] |
| 64 | $e + O_2^*(b) \rightarrow 2O + e$ | $6.86\times10^{-15}\times\exp(-4.66/T_e)$ | [29] |
| 65 | $e + O_2^*(b) \rightarrow O_2 + e$ | $9.72\times10^{-16}\times\exp(-0.591/T_e)$ | [29] |
| 66 | $O^- + O_2^*(b) \rightarrow O_2 + O + e$ | $6.9\times10^{-16}$ | [29] |
| 67 | $O + O_2^*(b) \rightarrow O + O_2^*(a)$ | $8.1\times10^{-20}$ | [29] |
| 68 | $O + O_2^*(b) \rightarrow O + O_2$ | $4\times10^{-20}$ | [34] |
| 69 | $O_2 + O_2^*(b) \rightarrow O_2 + O_2^*(a)$ | $3.79\times10^{-22}\times(300/T_g)^{-2.4}\times\exp(-281/T_g)$ | [29] |
| 70 | $O_2^*(b) + Ar^+ \rightarrow O_2^+ + Ar$ | $1.1\times10^{-16}$ | [44] |
| 71 | $O_2^*(b) + O_3 \rightarrow O_2 + O_2 + O$ | $1.5\times10^{-17}$ | [29] |
| 72 | $e + O_2^H \rightarrow O + O^* + e$ | $3.49\times10^{-14}\times\exp(-1.42/T_e)$ | [29] |
| 73 | $e + O_2 \rightarrow O_2^H + e$ | $1.13\times10^{-15}\times\exp(-3.94/T_e)$ | [29] |
| 74 | $e + O_2^*(b) \rightarrow O_2^H + e$ | $1.13\times10^{-15}\times\exp(-2.31/T_e)$ | [29] |
| 75 | $e + O_2^*(a) \rightarrow O_2^H + e$ | $1.13\times10^{-15}\times\exp(-2.96/T_e)$ | [29] |
| 76 | $O + O_2^H \rightarrow O + O_2$ | $4.95\times10^{-18}$ | [29] |
| 77 | $O + O_2^H \rightarrow O^* + O_2^*(b)$ | $1.35\times10^{-18}$ | [29] |
| 78 | $O + O_2^H \rightarrow O^* + O_2^*(a)$ | $2.7\times10^{-18}$ | [29] |
| 79 | $O_2 + O_2^H \rightarrow 2O_2^*(b)$ | $2.9\times10^{-19}$ | [29] |
| 80 | $e + O_2^H \rightarrow O + O^-$ | $7.32\times10^{-16}\times T_e^{-1.072}\times\exp(-0.468/T_e)$ | [29] |

| #  | Reactions | Rate coefficient | Reference |
|----|-----------|------------------|-----------|
| 81 | $e + O_2^H \rightarrow 2O + e$ | $6.86 \times 10^{-15} \times \exp(-1.79/T_e)$ | [29] |
| 82 | $O_2^- + O_2^*(a) \rightarrow e + O_2 + O_2$ | $2 \times 10^{-16}$ | [29] |
| 83 | $O_2^- + O_2^*(b) \rightarrow e + O_2 + O_2$ | $3.6 \times 10^{-16}$ | [29] |
| 84 | $O_2^- + O \rightarrow O_2 + O^-$ | $1.755 \times 10^{-16}$ | [34] |
| 85 | $O_2^- + O \rightarrow O_3 + e$ | $2.145 \times 10^{-16}$ | [29] |
| 86 | $O_2^- + O_2^+ \rightarrow O_2 + O_2$ | $2 \times 10^{-14} \times (T_g/300)^{0.5}$ | [29] |
| 87 | $O_2^- + O_2^+ \rightarrow 2O + O_2$ | $2 \times 10^{-14} \times (T_g/300)^{0.5}$ | [29] |
| 88 | $O_2^- + O^+ \rightarrow O + O_2$ | $2 \times 10^{-14} \times (T_g/300)^{0.5}$ | [29] |
| 89 | $O^- + O_3 \rightarrow O_2 + O_2^-$ | $0.3 \times 10^{-15}$ | [29] |
| 90 | $O^- + O_2^*(a) \rightarrow O + O_2^-$ | $4.75 \times 10^{-17}$ | [29] |
| 91 | $e + O_3 \rightarrow O + O_2^-$ | $9.76 \times 10^{-14} \times T_e^{-1.309} \times \exp(-1.007/T_e)$ | [29] |
| 92 | $O^- + O_2 \rightarrow O + O_2^-$ | $4.75 \times 10^{-18}$ | [45] |

Note: (1) $^a$ denotes the integral of the electron collision cross section obeying the Maxwell electron energy distribution. (2) $T_e$ is the electron temperature in eV. $T_g$ is the background gas temperature in K. (3) $O_2^*(a)$, $O_2^*(b)$, $O_2^H$ indicate $O_2(a^1\Delta g)$, $O_2(b^1\Sigma g^+)$, $O_2(A^3\Sigma_u^+, A^3\Delta_u, c^1\Sigma_u^-)$.

Table A2 Surface reactions, wall quenching and wall recombination coefficients.

| #  | Reactions | γ | Reference |
|----|-----------|---|-----------|
| 93 | $O_2^*(a) \rightarrow O_2$ | 0.007 | [34] |
| 94 | $O_2^*(b) \rightarrow O_2$ | 0.1 | [34] |
| 95 | $O_2^H \rightarrow O_2$ | 0.1 | [34] |
| 96 | $O \rightarrow \frac{1}{2}O_2$ | $0.1438 \times \exp(2.5069/p)$ | [34] |
| 97 | $O^* \rightarrow \frac{1}{2}O_2$ | $0.1438 \times \exp(2.5069/p)$ | [34] |
| 98 | $O^* \rightarrow O$ | 0.1 | [34] |
| 99 | $Ar^* \rightarrow Ar$ | 0.1 | [29] |
| 100 | $Ar^+ \to Ar$ | $u_{B,Ar^+} \dfrac{A_{eff}}{V}$ | [29] |
| 101 | $O_2^+ \to O_2$ | $u_{B,O_2^+} \dfrac{A_{eff}}{V}$ | [29] |
| 102 | $O^+ \to O$ | $u_{B,O^+} \dfrac{A_{eff}}{V}$ | [29] |

Note: (1) p denotes pressure in mTorr. (2) $O_2^*(a)$, $O_2^*(b)$, $O_2^H$ indicate $O_2(a^1\Delta g)$, $O_2(b^1\Sigma g^+)$, $O_2(A^3\Sigma_u^+, A^3\Delta_u, c^1\Sigma_u^-)$.